\documentclass[preprintnumbers,twocolumn,nofootinbib]{revtex4}
\usepackage{graphicx}
\usepackage{dcolumn}
\usepackage{bm}
\usepackage{epsfig,amsmath}
\usepackage{amssymb}
\usepackage{subfigure}
\usepackage[usenames,dvipsnames]{color}
\usepackage[pagebackref=false, colorlinks=true]{hyperref}
\definecolor{redish}{rgb}{0.7,0.2,0.0}  
\definecolor{bluish}{rgb}{0.2,0.5,0.8}

\hypersetup{linkcolor=redish,          
                  citecolor=blue,        
                  filecolor=magenta,      
                  urlcolor=bluish}          

\DeclareFontFamily{U}{rsfs}{}         
\DeclareFontShape{U}{rsfs}{m}{n}{<5> rsfs5 <6><7> rsfs7          %
  <8><9><10><10.95><12><14.4><17.28><20.74><24.88> rsfs10}{}     %
\DeclareMathAlphabet{\mathfs}{U}{rsfs}{m}{n}

\def \f{\frac}

\def \O{\Omega}

\def \L{\Lambda}

\def \th{\theta}

\begin{document}

\title{Black holes shielded by magnetic fields}

\author{Chandrachur Chakraborty}
\email{chandrachur.c@manipal.edu} 
\affiliation{Manipal Centre for Natural Sciences, Manipal Academy of Higher Education, Manipal 576104, India}

\begin{abstract}
Black holes (BHs) formed by collapsing and/or merging of magnetized progenitors, have magnetic fields penetrating the event horizon, and there are several possible scenarios. Thus, the no-hair theorem that assumes the outside medium is a vacuum, is not applicable in this case. Bearing this in mind and considering a Schwarzschild BH of mass $M$ immersed in a uniform magnetic field $B$, we show that all three frequencies related to the equatorial circular orbit of a test particle become imaginary for the orbits of radii $r_B > 2B^{-1}$. It signifies that if a BH is surrounded by a magnetic field of order $B \sim R_g^{-1}$ (where $R_g$ is the gravitational radius of the BH), a test particle could unable to continue its regular geodesic motion from/at $r > r_B$, hence the accretion disk could not be formed, and the motion of other stellar objects around the BH could be absent. As the BHs are generally detected by watching for their effects on nearby stars and gas, a magnetic field of order $B \sim R_g^{-1}$ could be able to shield a BH in such a way that it could remain undetectable. Motivated with this theoretical investigation and considering the sphere (of radius $r_f$) of magnetic influence around an astrophysical BH, we constrain $B$, above which a magnetized BH could remain undetectable. For example,  $M=10^9M_{\odot}$ BH surrounded by $B > 10^6$ G and $M=10M_{\odot}$ BH surrounded by $B > 10^{14}$ G could remain undetectable for $r_f \sim 10^5R_g$. In other words, our result also explains why a {\it detected} SMBH has surprisingly weak magnetic field.
\end{abstract}

\maketitle

\section{Introduction}
The Event Horizon Telescope (EHT) has recently imaged the polarized emission around the supermassive black hole (SMBH) M87* on event-horizon scales, and estimated the average magnetic field $\sim 1-30$ G \cite{eht7}. In an another interesting paper \cite{eatough}, it is mentioned that the emission models of Sagittarius A* require magnetic fields of about $30-100$ G to explain the synchrotron radiation from close to the event horizon. It is also inferred that the upper limit of the magnetic field in the X-ray corona of the BH Cygnus X-1 rises to $10^7$ G \cite{santo}. These are just a few example. Almost all the BHs are perhaps surrounded by a non-zero magnetic field, even it is very less compared to the magnetic field  ($10^{15}$ G) of a magnetar (the origin and the characteristics of which are also completely different from a BH). In general, the Gauss order magnetic field is considered as strong enough in case of a (supermassive-)BH. Although the BHs formed by collapsing of magnetized progenitors, have magnetic fields penetrating the event horizon, and there are several possible scenarios, we have not yet found such a BH in our Universe, which possesses a magnetic field of the same order of magnitude of a magnetar.

The motion of {\it charged} particles has been widely studied \cite{fs, shi, za, za2} considering a weakly magnetized Schwarzschild BH, and many interesting result found. On the other hand, the motion of a {\it neutral} test particle (which is very important from the astrophysical point of view) in the strongly magnetized Schwarzschild BH is not studied so far. It might be due to the following reason that a strong magnetic field has not been detected around a BH yet and/or it is generally assumed that a strong magnetic field cannot exist around a BH. 
A relatively more plausible physical reason for the absence of detection of strong magnetic fields associated with BHs could be due to the vacuum breakdown which allows the vacuum to break down for the creation of electron-positron pairs \cite{bz}. This argues against the accumulation of dynamically significant magnetic fields onto a BH, and the same remain undetected.
However, considering the simple enough magnetized BH spacetime (i.e., magnetized Schwarzschild) with no assumption on the magnetic field intensity, here we show that a strongly magnetized BH might exist in nature but remain undetectable.

The BHs are generally detected by watching for their effects on nearby stars (which orbit the BH),  and gas (which forms the accretion disk around a BH). In both the cases, the motion of the test particles is governed by the Kepler motion. An orbiting test particle can have three different frequencies due to its motion, which play a significant role in astrophysics. The Kepler frequency gives the rate of rotation of a test particle in an orbit, the periastron precession frequency gives the precession of the orbit and the orbital/nodal plane precession frequency gives the precession of the orbital plane. The true nature of the central object could be inferred \cite{ckp, cbgm} by analyzing these three frequencies (for example, see \cite{bs, sv}). Studying the nature of these three fundamental frequencies in the magnetized Schwarzschild spacetime, we constrain the magnetic field for a astrophysical BH, above which a magnetized BH could remain undetectable.

One should note here the difference between the accretion in quasars and that in low-luminosity active galactic nuclei. The former is well-modeled by the Novikov-Thorne like \cite{nt} thin-disk accretion, where the particles move on the circular timelike Keplerian orbits down to the innermost-stable-circular-orbit (ISCO). For the latter, the hot accretion flows extend all the way down to the horizon and tend to be sub-Keplerian. However, the arguments presented in this current work can also be applied approximately to the case of hot accretion. In fact, in this letter we have sometimes mentioned the accretion for M87* and Sgr A* as examples. The reason behind that is, both objects are quite well-known due to the recent EHT observations \cite{eht7, ehtsgr} which also establish the presence of horizon-scale magnetic fields for M87* \cite{eht7}.

The BHs formed by collapsing and/or merging of the magnetized progenitors, have magnetic fields penetrating the event horizon \cite{lyu83}. Thus, the no-hair theorem that assumes the outside medium is a vacuum, is not applicable in this case \cite{lyu83}. 
However, the gravitational collapse induced by an endoparasitic BH through capturing of dark matter into a neutron star (NS) \cite{gold}, spinning down of a rotationally supported hypermassive NS \cite{fal}, by acquiring the magnetic flux through merging of a BH with one (or multiple) magnetized NS(s) \cite{east} or magnetar(s), or by the accretion mechanism, a BH is magnetized. 

In fact, in case of a BH formation by collapsing of a NS, the magnetic field lines are effectively ``frozen in'' the star both before and during collapse \cite{lyu}. In the limit of no-resistivity \cite{lyu}, this introduces a topological constraint that prevents the magnetic field from sliding off the horizon \cite{lyu83}. Consequently, the newly-formed BH conserves the magnetic flux during collapse of a magnetized NS/magnetar to a BH. Now, if a non-rotating BH is surrounded by a uniform magnetic field, it could be described by the magnetized Schwarzschild BH metric which we are going to discuss in the next section.

\section{Magnetized Schwarzschild black hole}
\begin{figure}
 \begin{center}
 {\includegraphics[width=2.8in,angle=0]{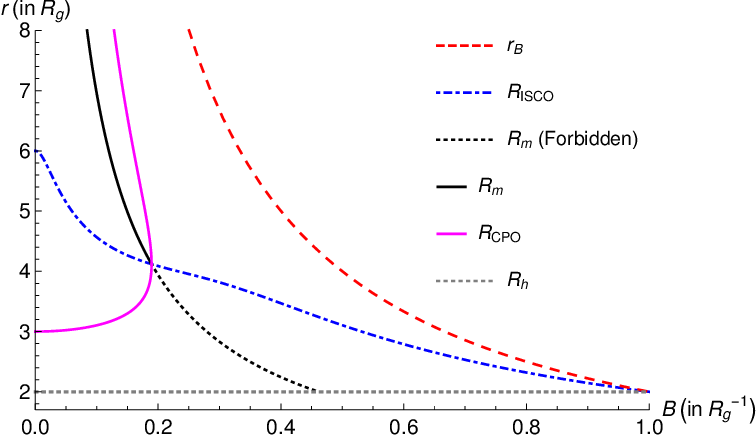}}
\caption{\label{newf}Various values of $r$ (in `$R_g$') vs $B$ (in `$R_g^{-1}$'). Although $r_B$ and $R_{\rm ISCO}$ coincide with $R_h$ for $B=R_g^{-1}$, both of them disappear behind the horizon for $B > R_g^{-1}$. $R_m$ coincides with $R_{\rm ISCO}$ and $R_{\rm CPO}$ at $r=4.12R_g$ for $B=B_{\rm cr}$. }
\end{center}
\end{figure}

The exact electrovacuum solution (in the geometrized unit) of the Einstein-Maxwell equation for the  magnetized Schwarzschild BH of mass $M$ can be written as \cite{er}
\begin{eqnarray}\nonumber
 ds^2 &=& \L^2\left[- \left(1-\f{2M}{r} \right)dt^2+\f{dr^2}{1-\f{2M}{r} }+ r^2 d\th^2 \right]
 \\
 &+& \L^{-2}r^2\sin^2 \th d\phi^2
 \label{sz}
\end{eqnarray}
where $\L=1+\f{1}{4}B^2r^2\sin^2\th$,
and $B$ is a constant and homogeneous magnetic field  around the BH, directed along the polar axis of a spherical coordinate system \cite{gp}. If a non-rotating \footnote{Note that the Kerr BH immersed in a uniform magnetic field was  considered upto the linear order in $B$ in \cite{Wald74a} (see Eq. 3.2 of \cite{Wald74a}). Thus, the background metric (Eq. 3.2 of \cite{Wald74a}) remained as the pure Kerr metric and was not a function of $B$ as well. For this reason, the same is not applicable to the strong magnetic field like $B \sim R_g^{-1}$.
On the other hand, in our case, Eq. (\ref{sz}) is the exact electrovacuum solution (non-rotating) of the Einstein-Maxwell equation, and the metric (Eq. \ref{sz}) is a function of $B$. No approximation to $B$ is considered in Eq. (\ref{sz}).} BH immersed in a magnetic field $B$, the above-mentioned metric (Eq. \ref{sz}) can describe its surroundings. Now, by virtue of the cylindrical symmetry, one needs to compare the energy ($\pi r^2 R_gB^2/8\pi$) of the magnetic field for a cylinder of radius $r$ and \footnote{$G_N$ is the Newton's gravitational constant and $c$ is the speed of light in the vacuum.} height $R_g$ ($\equiv G_NM/c^2$), with the mass $M$ itself \cite{gp}, to
estimate the relative contribution of the magnetic field
energy and mass $M$ of the BH to the metric.
Gal'tsov and Petukhov \cite{gp} found then that a homogeneous magnetic field starts to distort the metric at a distance from $r=0$ in the order of $B^{-1}$. From this, they obtained a characteristic gravitational scale of the magnetic field strength close to a BH of mass $M$ as \cite{gp, ag2}
\begin{eqnarray}
 B \simeq B_{\rm M} &\sim & 2.4 \times 10^{19} \f{M_{\odot}}{M} ~{\rm Gauss}
 \label{bm}
 \\
 &=& 2.4 \times 10^{19} \left(\f{G_NM_{\odot}/c^2}{R_g}\right) ~{\rm Gauss} \nonumber
\end{eqnarray}
where $M_{\odot}$ is the solar mass. Note that as an electrovacuum solution of the Einstein-Maxwell equation, the scalar curvature or the Ricci scalar is zero for Eq. (\ref{sz}), whereas the Kretschmann scalar (see Eq. (A1) of \cite{cmglp}) is a function of $B$ and $M$. Actually, the stronger the magnetic field is, the more it is concentrated near the axis of symmetry, i.e., $\th=0$ and $\th=\pi$ \cite{gp}. This is signified by the value of Larmor radius: $r_B=2B^{-1}$ \cite{brp} along the equatorial plane $\th=\pi/2$ (see the red dashed curve of FIG. \ref{newf} and the orange dotted curve of FIG. 1 of \cite{sup}). 

\begin{figure}
 \begin{center}
 {\includegraphics[width=2.8in,angle=0]{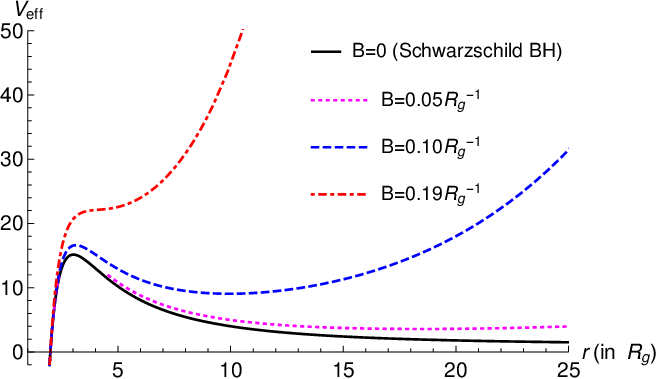}}
\caption{\label{vfig}$V_{\rm eff}$ vs $r$ for various values of $B$ for $l=20R_g$.  For $B > B_{\rm cr}=0.189R_g^{-1}$, the arbitrarily high effective potential does not allow a test particle to cross the barrier (indicated by the red dot-dashed curve) located at $r_B \approx 10.53R_g$.}
\end{center}
\end{figure}

As $B$ is expressed in $R_g^{-1}$ unit, $r_B$ is in the length unit (i.e., $R_g$). Here, one should also note that the geometrized unit of $B$ is meter$^{-1}$, and $1$ m$^{-1}$ in the geometrized unit is equivalent to $1.23 \times 10^{23}$ G. Although a magnetic field of order $B_M$ significantly distorts the spacetime, it does not change the form of the radius ($R_h$) of the horizon and the so-called surface gravity \cite{gp}. One intriguing feature is that, even a sufficiently small magnetic field $B << B_M$ that extends to a sufficiently large distance, begins to influence the metric at large distance $r$ \cite{gp}, in principle.  
From the theoretical point of view, there exist a critical value of magectic field: $B_{cr} =2\sqrt{3}(169+38\sqrt{19})^{-1/2}R_g^{-1} \approx 0.189R_g^{-1}$ above which no circular photon orbit (CPO) exists \cite{gp} (see FIG. \ref{newf} and the related discussion in \cite{sup}). In addition, the arbitrarily high effective potential ($V_{\rm eff}$) for $B \geq B_{\rm cr}$ does not allow a test particle (even a photon) to cross the barrier of $V_{\rm eff}$ (see FIG. \ref{vfig} and the related discussion in \cite{sup}).
Thus, from the theoretical point of view, the upper limit of $B$ is $B = 0.189R_g^{-1}$, above which ($B > 0.189R_g^{-1}$) a BH should remain undetectable.

The magnetized Schwarzschild metric (Eq. \ref{sz}) is not asymptotically flat, which means that the magnetic field is extended to infinity.
However, in reality, the magnetic field of a BH cannot be extended upto $r \rightarrow \infty$. Therefore, for the practical purpose, here we consider the uniform magnetic fields are perceived upto a fixed radius $r_f$ (say, $r_f \sim 10^5R_g$ \cite{fkr, bp}) \footnote{\label{fnt}Although the sphere of gravitational influence could be extended upto $r \sim 10^7R_g$, the uniform $B$ is not necessarily extended upto the same distance. $r \sim 10^2R_g$  \cite{bp, cb17} is considered to be close to a BH in the accretion physics \cite{bcb2}, whereas the outer radius of the disk is generally considered as $\sim 10^5R_g- 10^6R_g$ \cite{fkr, bp, bcb19, bcb2}. In case of the orbiting star around a BH, the reference of the farthest (closest) star S85 (S4711) orbiting Sgr A* with a period $P=3580$ yr ($7.6$ yr \cite{pei}) with the orbital radius $\sim 10^6R_g$ ($\sim 10^4R_g$) could be also taken into consideration. Here we consider that the  stars and gas orbit a BH safely (without being destroyed by the central BH instantaneously), if their orbits occur in the range of $10^4R_g-10^6R_g$. In that case, they should be fairly detected from the astrophysical observation. Referring to the above scenarios, here we consider the sphere of magnetic influence  as $r_f \sim 10^5R_g$. This astrophysically detectable range should be valid for both the accretion disk and orbiting stars. Using this idea, we can constrain $B$, above which a magnetized BH could remain undetectable.} around an astrophysical BH. The similar structure is also considered in other papers \cite{bamba, eatough, cy} as well. For example, the typical size of the uniform magnetic field ($B \sim 10^{15}$ G) for a magnetar is taken as $10^8$ km \cite{bamba} which is $10^7$ times of its radius (i.e., $\sim 10$ km). In our case, we have studied those BHs which are formed by collapsing and/or by merging of the magnetized progenitors (e.g. neutron stars, magnetars), and that the magnetic field cannot `slide off' from these BHs as shown in \cite{lyu83}. Therefore, considering all the above-mentioned facts, we choose $r_f$ as $10^5$ times of the radius ($\sim 2R_g$) of the BH. Note that $r_f \sim 10^5R_g$ is not considered as a stringent upper/lower limit. Rather, it ($r_f \sim 10^5R_g$) is chosen as an empirical value for the purpose of demonstration in a realistic scenario, as discussed above.

Thus, $r_B$ does not arise at all in reality for $B << B_M$, as $r_B >> r_f$. For example, $B=10^2$ G$=10^{-21}$ m$^{-1}$ \cite{eatough} seems to be extremely weak compared to $B_M \sim 10^{12}$ G $\sim 10^{-11}$ m$^{-1}$, if one fits/models the supermassive BH (SMBH) Sgr A* in the magnetized Schwarzschild metric mentioned in Eq. (\ref{sz}). Notice, here the calculated $r_B \sim 10^{21}$ m $\equiv 10^{11}R_g >> r_f$  is unfeasible, hence $r_B$ does not occur in this case. For $r_B >> r_f$ (applicable only for the weakly magnetized Schwarzschild BH), the effect of magnetic fields at $r > r_f$ is almost zero.

\section{Magnetically shielded black holes}
It was pointed out in \cite{gp} that the magnetic field does not only produce a barrier for the plasma, but also for the neutral particles and prevents their escaping to the infinity.  These particles create an atmosphere near the event horizon of the BH having the shape of a surface of revolution \cite{gp}. A magnetic field around the BH with a specific radius $r_f$ (sphere of magnetic influence) could be now perceived due to this. In such a situation, $r_B$ plays an important role, i.e., if $r_B \leq r_f$, the magnetic field at $r < r_f$ might be able to shield the BH from the {\it external disturbances}. 

Now, as all the three frequencies (see Eqs. 3 -- 5 of \cite{sup}) become imaginary \cite{cmglp} for $r > r_B$, a test particle motion in a circular geodesic starting from/at $r \geq r_B$ seems to be unfeasible, provided $r_B$ occurs for this specific case, i.e., $r_B \leq r_f$. If the value of magnetic field is much lower than $ R_g^{-1}$, i.e., $B << R_g^{-1}$, a test particle motion along the geodesic is not hampered in the realistic cases. In that case, $r_B$ is much bigger than $R_h$ and $r_f$ as discussed before, and, hence, $r_B$ could not arise at all in a realistic case. Due to the same reason, Sgr A* should not face any problem to accrete matter from other astronomical objects. There is also a possibility of disruption of the axisymmetric accretion disk (known as magnetically arrested disk) in presence of a {\it strong} magnetic field (not as strong as $B \sim 0.189R_g^{-1}$), due to which the matter eventually free-falls following the magnetic field lines \cite{nia} depending on the other parameter values and specific conditions.

If a strong magnetic field in the order of $B \sim 0.189R_g^{-1}$ or slightly below of it exists around a Schwarzschild BH, a surface of revolution \cite{gp} near the event horizon could be formed and $r_B$ occurs within $r_f$ (i.e, $r_f \gtrsim r_B$). As the regular geodesic motion is not possible for $r \geq r_B$, the BH cannot accrete matter from the other objects those exist outside of this barrier. In spite of that, if an object (e.g., stars, small BHs etc.) comes at $r < r_B$, the accretion disk should be formed, in principle. But the wideness of the disk, i.e., $(r_B-R_{\rm ISCO})$ is too narrow to form an accretion disk for $B \sim 0.189R_g^{-1}$, where $R_{\rm ISCO}$ is the radius of ISCO (see Eq. 8 of \cite{sup} and FIG. \ref{newf}). Moreover, the particles of the disk should not follow the geodesics due to the presence of strong magnetic fields, as shown in \cite{nia}. In fact, it is hardly possible to cross the surface of revolution containing charged and neutral particles \cite{gp} and reach the horizon (see FIG. \ref{vfig}). Eventually, for the existence of such a strong magnetic field, the accretion disk could not be formed in reality. The astronomers generally detect the BHs by watching for their effects on nearby stars and gas. If the accretion disk do not form and the regular geodesic motion is absent, it would be extremely difficult to detect such a  BH which is immersed in a strong magnetic field ($B \sim R_g^{-1}$). Even CPO does not occur for $B > 0.189R_g^{-1}$ \cite{gp}, as seen from FIG. \ref{newf}. In fact, the BHs surrounded by $B > 0.189R_g^{-1}$ cannot be detected by any means, as all the test particle motion including the photons motions are refrained to cross the arbitrarily high potential barrier (see FIGs. \ref{newf} and \ref{vfig}). We can call them as the magnetically shielded black holes. 

\begin{figure}[h!]
 \begin{center}
 {\includegraphics[width=2.8in,angle=0]{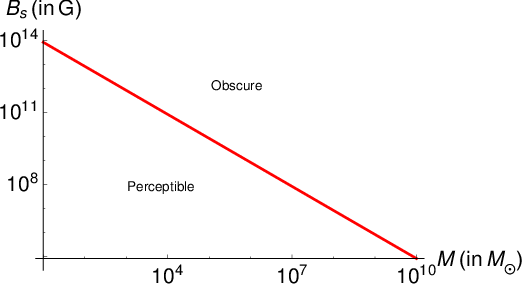}}
\caption{\label{f3}The solid red line represents the upper limit of magnetic field $B_s=8.2 \times 10^{14}M_{\odot}/M$ G for detection of a magnetized BH, which divides the $B_s-M$ space in the perceptible and obscure regions. For example, a $M=10^9M_{\odot}$ BH might remain undetectable or obscure if it is surrounded by $B \gtrsim 10^{6}$ G.}
\end{center}
\end{figure}

For a realistic astrophysical scenario, a BH with much lower value of $B_{cr}$ could remain undetectable. For example, if $r_B$ occurs within the range of $r_f$ (see the discussion of footnote \ref{fnt}), the upper limit of $B$ is calculated as $B_s \sim 1/r_f$ instead of $B=2/r_B$. Considering \footnote{Motivated with our theoretical result, we try to show that a magnetic field could shield a BH from the detection in reality, and, hence, we choose (by some physical arguments) an empirical value $r_f \sim 10^5R_g$ for the realistic scenario. If one choose $r_f \sim 10^3R_g$ (or some other values instead of $r_f \sim 10^5R_g$), the values of $Y$ axis of FIG. \ref{f3} will be increased by the two orders of magnitude as shown Eq. (\ref{eq3}).} $r_f \sim 10^5R_g$, one obtains 
\begin{eqnarray}
 B_s \sim 10^{-5}R_g^{-1}=8.2 \times 10^{14}M_{\odot}/M \,\,\,\, {\rm Gauss}.
 \label{eq3}
\end{eqnarray}
Therefore, if a BH of mass $M$ is surrounded by $B > B_s$, its existence could be obscure. On the other hand, the BHs with $B < B_s$ are perceptible. This is depicted in FIG. \ref{f3} .

While this type of isolated BHs do not produce detectable emission, their gravity can bend and focus light from background objects. Therefore, there may a possibility to detect them with the help of gravitational lensing, but the measurement of magnetic field intensity $B$ around the same might remain unsuccessful from the present technological expertise. Therefore, such type of highly magnetized BHs remain imperceptible to us.

\section{Conclusion and Discussion}
Considering the simple enough magnetized BH metric, we have first theoretically shown that it is impossible to detect a BH, if it is immersed in a magnetic field of $B > B_{\rm cr} \approx 0.189R_g^{-1}$. With the absence of any orbital motion of a test particle, the photon motion is also seized and CPO does not occur for $B > B_{\rm cr}$. Motivated with this theoretical result, we have constrained the magnetic field for the astrophysical BHs, in the second part of this letter. We have shown in FIG. \ref{f3} the intensity of magnetic field $B_s$ around a BH, above which ($B_s > 10^{-5}R_g^{-1}$) it could not be detected from the astrophysical observation. For example, a $M=10^9M_{\odot}$ and a $M=10M_{\odot}$ BHs could remain undetected if they are surrounded by $B \sim 10^6$ G  and $B \sim 10^{14}$ G, respectively. This is because of the presence of $r_B$ which is around $r_f \sim 10^{5}R_g$. In other words, our result explains why a {\it detected} SMBH has surprisingly weak magnetic field. It also means that the strongly magnetized SMBHs may exist in nature, but it would remain undetectable. 

Note that our result is based on the similar assumptions made in \cite{lyu83}, and the no-hair theorem is not applicable in this case. If one still considers that the fate of the magnetic flux on the event horizon has to be in accordance with the no-hair theorem, the {\it only} way to lose the magnetic field of BH is to change its topology (reconnect), as proposed in
\cite{brp}. However, it is not necessarily true that the collapse of a magnetized progenitors naturally creates conditions favorable for efficient reconnection, as clearly argued in \cite{lyu83}. Therefore, the scenarios presented in \cite{lyu83} and \cite{brp} are not the same, and our proposition does not violate what presented in \cite{brp}.

 Secondly, we have studied here the trajectories of a single neutral test particle in a uniform magnetic field in a simplest-possible magnetized BH. This is similar to the case, as one deduces the three fundamental frequencies (e.g., Kepler frequency, radial and vertical epicyclic frequencies) \cite{sv, bs, ing, ckp, cbgm} in the accretion disk theory, assuming the motion of a test particle around a BH. We do not make our study complicated by introducing the plasma dynamics (following \cite{gp, shi, za, za2} and so on) as we think our assumption suffices for our purpose. One may definitely consider this problem in a complex scenario in a later stage. Our result could be improved if the more and more complex situation is considered, e.g. by introducing the plasma dynamics, orbital eccentricity, variation of the magnetic field and so on.
 
 The metric dealt with in this letter is not asymptotically flat, as the Schwarzschild BH is considered to be immersed in a uniform magnetic field. One plus point of the magnetized Schwarzschild metric is that it is the exact electrovacuum solution of Einstein-Maxwell equation. This is also a standard practice to consider a uniform magnetic field around the BH to explain the many astrophysical phenomena (e.g. see \cite{fs, shi, za, za2, tur, nia, dad, say} and so on), as it is easier to handle. Due to the same reason, here we have also considered the uniform magnetic field to find the upper limit of $B$ around a BH, for which the astrophysical detection of such a BH is possible. Note that here we have assumed that the BH maintains an approximately uniform magnetic field upto $r_f$. For a varying magnetic field, the value of $B_s$ is not supposed to change enormously in the region $r < r_f \sim 10^5R_g$. Considering the full (magnetized) Kerr metric \cite{cprd1}, we have already studied the influence of the Kerr parameter on $B_s$, but no significant changes are found on the same. We suppose to report it in a future publication.
\\

 {\bf Acknowledgements :} I thank the referee for constructive comments that helped to improve the manuscript. I dedicate this Letter to my mother Shree Snigdha Chakraborty who was my constant inspiration in all fields of life, and unfortunately passed away on September 11, 2023.

\onecolumngrid
\newpage
\begin{center}
\textbf{\large Supplementary Material: Black holes shielded by magnetic fields}
\vspace{.5cm}

Chandrachur Chakraborty
\\
\vspace{.1cm}
{\it Manipal Centre for Natural Sciences, Manipal Academy of Higher Education, Manipal 576104, India}
\end{center}
\setcounter{equation}{0}
\setcounter{figure}{0}
\setcounter{table}{0}
\setcounter{page}{1}

\vspace{1cm}

\section*{\label{scpo}Circular photon orbit in the magnetized Schwarzschild spacetime} ~

The mertic for the magnetized Schwarzschild black hole (BH) is introduced in Eq. (1) of \cite{ol}. The said metric describes a BH of mass $M$ surrounded by an uniform magnetic field $B$. The gravitational radius ($R_g$) of the BH
can be written as $R_g=G_NM/c^2$, where $G_N$ is the Newton's gravitational constant and $c$ is the speed of light in the vacuum. Deriving the circular photon orbit (CPO) equation:
\begin{eqnarray}
 3 B^2r^3-5R_gB^2r^2-4r +12R_g= 0
 \label{eqcpo}
\end{eqnarray}
for the above-mentioned BH, and solving it, we obtain two positive real roots for all $B$ values for $0 < B < B_{cr}$, where
\begin{eqnarray}
B_{cr} =2\sqrt{3}(169+38\sqrt{19})^{-1/2}R_g^{-1} \approx 0.189R_g^{-1}
\end{eqnarray}
is a critical value of magnetic field
above which no circular photon orbit (CPO) exists
\cite{gp}. Note that one obtains only one positive real root for $B=B_{cr}$. CPO does not occur for $B > B_{cr}$, as seen from the solid magenta curve of FIG. 1 of \cite{ol}. For $B \rightarrow 0$, the radii ($r_{\rm CPO}$) of the two CPOs are $3R_g$ and $\infty$ \cite{gp}, respectively.

\section*{Fundamental frequencies in the magnetized Schwarzschild spacetime\label{s3}}
\begin{figure}[h!]
 \begin{center}
 {\includegraphics[width=2.8in,angle=0]{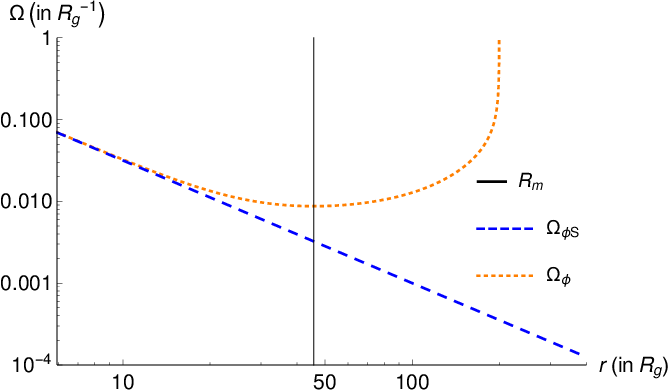}}
\caption{\label{f2}The Kepler frequency for magnetized ($\O_{\phi}$) and regular ($\O_{\phi S}$) Schwarzschild  BHs are drawn only for reference. The orange curve stands for $B=0.01R_g^{-1}$, and, hence, $R_{\rm ISCO}=5.92R_g, R_m=45.68R_g$ and $r_B=200R_g < r_f$.}
\end{center}
\end{figure}

The three fundamental frequencies related to the orbit of a test particle, and important for the accretion disk theory, are directly derived from the magnetic Schwarzschild metric components. These are, Keplerian frequency $\O_{\phi}$, periastron precession frequency : $\O_{\rm per}$, and orbital/nodal plane precession frequency: $\O_{\rm nod}$ \cite{ckp, cbgm}.  These three frequencies could be inferred by measuring the quasi-periodic-oscillations (QPOs) frequencies come from the inner accretion disk \cite{ckp, cbgm}. In a very recent paper \cite{cmglp}, the above mentioned three frequencies are obtained as
\begin{eqnarray}
 \O_{\phi} &=& \f{(4+B^2 r^2)^2}{16} \sqrt{\f{R_g(4-3B^2r^2)+2B^2r^3}{r^3(4-B^2r^2)}},
 \label{ope1}
\end{eqnarray}
\begin{eqnarray}
  \O_{\rm nod} &=& \O_{\phi}-\sqrt{\f{R_g}{r^3}},
\end{eqnarray}
and, 
 \begin{eqnarray}
 \O_{\rm per} &=& \O_{\phi}-\f{1}{r^2 (4 + B^2 r^2)}\sqrt{\f{N}{(4 - B^2 r^2)}} ,
 \label{ore1}
\end{eqnarray}
where 
\begin{eqnarray}\nonumber
 N &=& 4 B^2 r^4 (32 - 12 B^2 r^2 + 3 B^4 r^4) 
- R_g^2 (384 - 672 B^2 r^2 + 200 B^4 r^4 - 30 B^6 r^6) 
\\
&+& R_g r (64 - 624 B^2 r^2 + 204 B^4 r^4 - 37 B^6 r^6). 
\end{eqnarray}
A close observation reveals that all three frequencies (Eqs. \ref{ope1}--\ref{ore1}), become imaginary for $r > r_B=2B^{-1}$ which is depicted by the red dashed line in FIG. 1 of \cite{ol}. 
In general, all the above-mentioned three frequencies decrease to zero at $r \rightarrow \infty$ in any asymptotic spacetime, e.g., Schwarzchild, Kerr spacetimes etc., whereas they behave differently in the magnetized Schwarzschild spacetime. Setting 
$d \O_{\phi}/dr=0$ for $r=R_m$, one obtains 
\begin{eqnarray}
 12 B^6 R_m^7- 15 R_g B^6 R_m^6 - 80 B^4 R_m^5+ 132 R_g B^4 R_m^4 -208 R_g B^2 R_m^2 +192R_g=0.
 \label{rm}
\end{eqnarray}
Now, by solving Eq. (\ref{rm}) one can find the decreasing trend in the three frequencies (with increasing $r$) upto the fixed radius $r=R_m$ for the magnetized Schwarzschild spacetime, where $\O_{\phi}$ becomes minimum (for example, see FIG. \ref{f2}). This is expected in the asymptotic limit of any asymptotic spacetime. As Eq. (\ref{rm}) cannot be solved analytically, we solve it numerically and depicted by the solid black line in FIG. 1 of \cite{ol}.

\section*{Innermost stable circular orbit in the magnetized Schwarzschild BH\label{s3.1}}~
In a very recent paper \cite{cmglp}, the radius ($R_{\rm ISCO}$) of the innermost stable circular orbit (ISCO) is obtained by solving
\begin{eqnarray} \nonumber
&&12 B^6 r^8- 37 B^6 R_g r^7+ (30 B^6 R_g^2-48 B^4) r^6+204 B^4 R_g r^5+(128 B^2 - 200 B^4 R_g^2) r^4   \nonumber
\\
&-& 624 B^2 R_g r^3
+ 672 B^2 R_g^2 r^2+ 64 R_g (r-6R_g)=0
 \label{isco}
\end{eqnarray}
(one may also see Eq. 16 of \cite{cmglp} and the related discussion) at $r=R_{\rm ISCO}$. For a regular Schwarzschild BH (with $B=0$), $R_{\rm ISCO}=6R_g$ can be obtained from Eq. (\ref{isco}), whereas for $B=R_g^{-1}$ the ISCO coincides with the horizon $R_h=2R_g$. For $B > R_g^{-1}$, the ISCO occurs at $r < 2R_g$, in principle. Thus, all orbits at $r > 2R_g$ are stable in that sense. As the ISCO equation (Eq. \ref{isco}) cannot be solved analytically, we solve it numerically and obtain the dot-dashed blue curve as shown in FIG. 1 of \cite{ol}. If the value of $B$ increases from $0$ to $R_g^{-1}$ and more, the ISCO radius decreases from $6R_g$ to $2R_g$ and then disappears behind the horizon. In fact, for the realistic cases, it is too difficult to exist the stable circular orbits in such a high magnetic fields (e.g., $B \sim 10^{10}-10^{11}$ T) \cite{vrba}. For weak magnetic field, i.e., $B << R_g^{-1}$, the ISCO occurs close to $6R_g$ but the exact value is lower than $6R_g$ \cite{shy}. If we increase the value of $B$ from $0$ to $R_g^{-1}$, the difference between $(r_B-R_{\rm ISCO})$ decreases and vanishes on the horizon for $B=R_g^{-1}$. 
For further increment of $B$, i.e., $B > R_g^{-1}$, the all $r_B$ and $R_{\rm ISCO}$ are seem to be unfeasible.

\section*{\label{sveff}Effective potential}
The effective potential ($V_{\rm eff}$) for the magnetized Schwarzschild spacetime is written as \cite{gp}
\begin{eqnarray}
 V_{\rm eff}=\L_0^2 \left(1-\f{2R_g}{r} \right) \left(1+\f{l^2\L_0^2}{r^2} \right) 
 \label{veff}
\end{eqnarray}
where $l$ is the angular momentum of the test particle and $\L_0=1+r^2/r_B^2$. For $B \rightarrow 0$, $V_{\rm eff}$ resembles Eq. (9.28) of \cite{jh}. In the weak magnetic field (i.e., $B << R_g^{-1}$), the term $\L_0^2 \rightarrow 1$. On the other hand, $\L_0^2$ starts to dominate for $B \sim R_g^{-1}$, and make the potential barrier so high (see FIG. 2 of \cite{ol}) that the test particles cannot cross it at all.

\end{document}